\documentstyle[12pt,emulateapj]{article}
\input psfig.sty
\def\arcsec {$^{\prime \prime}$}
\def\arcmin {$^{\prime}$}

\def\etal   {{\it et~al.\/}}
\def\HI     {H{\sc I}}
\def\HII    {H{\sc {II}}}
\def\kms    {~km~s$^{-1}$}

\begin{document}

\title{Obtaining Galaxy Masses Using Stellar Absorption and 
[O~II] Emission-Line Diagnostics in Late-Type Galaxies}

\author{Henry A. Kobulnicky\altaffilmark{1}$^,$\altaffilmark{2}$^,$\altaffilmark{3}
and Karl Gebhardt$^{1,2}$}

\altaffiltext{1}{Hubble Fellow}
\altaffiltext{2}{Visiting 
Astronomer, Kitt Peak National Observatory, National Optical Astronomy  
Observatories, which is operated by the Association of Universities for 
Research in Astronomy, Inc. (AURA) under cooperative agreement with the 
National Science Foundation.  We are grateful to the legacy of 
KPNO as a national observatory, especially its support of fine
instruments on small telescopes to which we ascribe the success of 
this project.}
\altaffiltext{3}{Present address: Department of Astronomy, 
University of Wisconsin, 475 N. Charter St., Madison WI 53706}  

\affil{Lick Observatory \\ 
University of California, Santa Cruz \\ Santa Cruz, CA 95064 \\
Electronic Mail: chip@ucolick.org, gebhardt@ucolick.org}

\authoremail{chip@ucolick.org, gebhardt@ucolick.org}

\begin{abstract} 

The [O~II] $\lambda$3727 emission lines and absorption features from
stellar Balmer and Ca H\&K lines are the most accessible kinematic
diagnotics in galaxies at $z\sim1$.  We investigate the kinematics of
22 local late-type galaxies using these spectral features and we
compare the results to 21-cm neutral hydrogen spectra in order to
assess the utility of each diagnostic for measuring galaxy masses.  In
order to simulate data at high redshift where only 1-dimensional
velocity profiles are normally available, we study spatially integrated
as well as spatially-resolved spectra.  Although the studied galaxies
span a wide range of morphological types, inclinations, and star
formation rates, we find that the gaseous and stellar kinematic tracers
yield comparable kinematic line widths and systemic velocities.  The
[O~II] and H~I line widths correlate most strongly, showing an
intrinsic dispersion of $\sim$20 \kms, or $\sim10\%$ for a typical
galaxy with a kinematic width of 200 \kms.  In a few extreme cases, the
[O~II] line widths underestimate the neutral hydrogen width by 50\%.
Reliable velocity widths can also be obtained from the stellar Balmer
and Ca H\&K absorption lines, even for some of the very late--type
galaxies which have strong emission lines.  The intrinsic dispersion is
$\leq$10\% between the stellar absorption and H~I line widths.  We
provide a prescription for using these strong stellar
absorption and [O~II] emission features to measure the kinematics, and
thus masses, of galaxies in the distant universe.

\end{abstract}


\section{Introduction}

Kinematic studies of distant galaxies reveal their structure at early
times and help trace their evolution to the present day.   Measurements
of the most important kinematic properties, the velocity widths and
rotation curves, are now becoming routine for galaxies at
cosmologically significant distances.  Conclusions are mixed regarding
whether distant field galaxies are consistent with local Tully-Fisher
(T-F) relations out to $z=1$.  The results are consistent with modest
evolution in the T-F relation, in the sense that galaxies of a given
rotational amplitude are 0.2-1.5 magnitudes brighter at earlier epochs
(Forbes~\etal\ 1995; Vogt~\etal\ 1996, 1997; Simard \& Pritchet 1998;
Rix \etal\ 1997).  In these studies, the bluest galaxies deviate most
strongly from the T-F relation suggesting that the conclusions are
sensitive to target selection effects (Bershady~\etal\ 1998).  Since
current studies rely on the brightest and largest galaxies which have
well-defined rotation curves, the samples are small and may be biased
by including only the most massive spirals.    For early-type galaxies,
the velocity dispersions and luminosities provide an estimate of
evolution in the fundamental plane.  Recent work indicates luminosity
evolution for both cluster and field ellipticals.  Cluster ellipticals
show modest evolution---about 0.8 magnitudes of brightening for a fixed
kinematic width at z$\sim 0.8$ compared to $z=0$ (Kelson \etal\ 1997;
van Dokkum \& Franx 1996; van~Dokkum~\etal\ 1998; Treu~\etal\ 1999).
This evolution in the fundamental plane is consistent with very early,
single-burst, formation scenarios.  Field ellipticals show more
brightening, $\sim$1.5 magnitudes (Gebhardt~\etal\ 2000), suggesting a
fairly recent formation epoch.

Measuring the dynamics of galaxies at cosmologically significant
redshifts is observationally challenging because traditional kinematic
indicators are redshifted to wavelengths inaccessible from the ground
for many redshift regimes of interest.  For the early-type galaxies at
intermediate--to--high redshift, traditional stellar dynamical tracers
like H$\alpha$ emission, Mg~II $\lambda 5167,5173,5814$ and G-band
absorption features at $\lambda4300$, and the Ca~II
$\lambda8498,8542,8662$ triplet shift out of the optical window.  They
fall within near-IR atmospheric windows or between night sky lines only
for specific redshifts.  At redshifts of $z\simeq1$, the stellar Ca
H\&K and Balmer lines become the only absorption lines accessible with
optical spectrographs.  For late type galaxies, the [O~II] 3727
\AA\ doublet is the most promising emission line for kinematic
measurements.  It is important to investigate the utility of both
spectral features for measuring galaxy masses by comparing the [O~II]
velocity widths with those of other trusted emission lines (H$\alpha$
or 21-cm).  Early results are encouraging for the [O~II] widths.  There
is an excellent correlation between optical H$\alpha$ rotation curves
and 21-cm profiles (e.g.,  Mathewson, Ford, \& Buchhorn 1992;
Raychaudhury \etal\ 1997; Courteau 1997 and references therein).
Courteau \& Faber (2000) show a tight correlation between [O~II],
H$\alpha$, and \HI\ kinematics in a sample of Sc galaxies.

Another obstacle for mass estimates at intermediate $z$ is that most
galaxies are smaller than a typical 1\arcsec\ ground-based seeing disk,
resulting in unresolved rotation curves. For this reason, it is
important to investigate the effects of degraded spatial resolution on
kinematic measurements in distant galaxies by comparing global
1-dimensional (i.e., spatially integrated) galaxy spectra with
spatially-resolved spectra.

In this paper we present kinematic measurements for 22 local galaxies
covering a range of morphological types, from blue compact and
irregular, to spiral and active galaxies.  We explore whether
kinematics derived from [O~II] $\lambda\lambda$3726,3729 doublet
profiles yield the same results as H$\alpha$ and 21-cm neutral hydrogen
data. We also compare the velocity widths derived from global emission
line and 21-cm profiles with those derived from Balmer and Ca H\&K
lines to explore the utility of stellar absorption features for
measuring masses in late type galaxies where absorption features have
not traditionally been used.  We find good agreement between different
dynamical tracers.  In a small fraction of extreme objects, or in
data with low S/N, the [O~II] line widths may underestimate the kinematic
widths by up to 50\%.

\section{Spectroscopic Observations of Local Galaxies}

We obtained optical longslit spectroscopy of 22 local ($V<4000$ \kms)
galaxies over the wavelength range 3650 \AA\ - 4600 \AA\ with the Kitt
Peak National Observatory 2.1~m telescope and GoldCam spectrograph
during 1999 February 15--19.  Table~1 summarizes the targets, their
morphological types, and their basic physical properties.  Targets were
selected to include a diverse assortment of morphological types from
blue compact to Sb spiral galaxies.  We include 8 objects from the
nuclear starburst study of Lehnert \& Heckman (1995): NGC 2146,
NGC~2798, NGC2966, NGC~3044, NGC~3593, NGC~4527, NGC~4666, and
NGC~4818.  Most targets have absolute magnitudes in the range
$-18<M_B<-20.5$ with inclinations $30<i<75$.  Thus, the sample includes
objects similar to galaxies used for kinematic studies at
intermediate redshifts (Forbes~\etal\ 1995; Vogt~\etal\ 1996, 1997;
Simard \& Pritchet 1998; Rix \etal\ 1997).

The observed spectral range includes the [O~II] $\lambda$3727 emission
line to measure the kinematics of the ionized gas, and the stellar
Balmer and the Ca H\&K $\lambda\lambda$3970, 3932 lines to measure the
kinematics of the underlying stellar population.  Other prominent
features in this spectral range include H$\delta$ and H$\gamma$ which
may be seen either in emission from ionized gas or in absorption from
stellar atmospheres.  The KPNO grating \#47, used in second order with
a slitwidth of 2.5\arcsec\ and a CUSO$_4$ filter to block first order,
yielded a spectral dispersion of 0.47 \AA\ pix$^{-1}$ and a mean
spectral resolution of  1.57 \AA\ (126 \kms) FWHM at 3727 \AA\ (2.35
\AA\ $\equiv$156 \kms\ at 4500 \AA).  The spatial scale of the CCD was
0.78\arcsec\ pixel$^{-1}$, and seeing averaged 2\arcsec.  Calibration
included corrections for bias, flatfield, slit illumination and cosmic
ray rejection.  Frequent exposures of an HeNeAr arc lamp provided a
wavelength calibration accurate to 0.05 \AA\ RMS ($\sim$4 km/s).  The
spatial and spectral characteristics of the spectrograph remained
constant from night-to-night and from position to position on the sky,
varying by $<1$ pixel (0.47 \AA) throughout the run.  Furthermore, the
dispersion vector was aligned nearly perfectly with the CCD rows so
that the wavelength solution (dispersion and zero point) changed by
less than 1 pixel over the entire usable detector area.  Based on the
frequency and stability of the arc lamp exposures, we estimate that
relative radial velocities from exposure to exposure are accurate to
0.25 pixels (10 \kms).  All velocities reported here correspond to the
heliocentric reference frame.  Occasional cirrus clouds prohibited
photometric calibration, but relative fluxes based on
spectrophotometric standard stars and radial velocity calibrations
based on frequent arc lamp exposure were not affected.

For each galaxy we obtained total integration times of 40 to 60
minutes, broken into multiple exposures to aid cosmic ray rejection.
We oriented the 5\arcmin\ slit along the major axis of each galaxy at
the position angle listed in Table~1 and as illustrated in Figure~1.
For most of the objects, we also obtained several exposures while the
telescope drifted across the galaxy in a direction perpendicular to the
slit.  These exposures produce a ``global'' spectrum most closely
approximating the spectra being acquired of high-redshift galaxies with
typical ground-based resolution elements (several kpc).  Generally the
drift scans have lower signal-to-noise than the fixed exposures since a
smaller fraction of the integration is spent on the high surface
brightness portion of the galaxy.  Table~1 records the spatial
amplitude of the drift for applicable galaxies.  Images of the target
galaxies appear in Figure~1 along with markers showing the location of
the fixed slit exposures and the extent of the drift scan exposures.

We observed 12 stars with well-known spectral types from Leitherer
\etal\ (1996) and Lick IDS spectra from Worthey \etal\ (1994) in order
to construct templates for the measurement of stellar absorption
features in the program galaxies.  The template stars span a range of
spectral types, A2 through K2, and a range of luminosity classes, dwarf
through supergiant.  Table~2 lists these template stars.

Comparable exposures of the same galaxy were averaged to reject cosmic
rays and then background subtracted to remove night sky emission
lines.  Background subtraction was also necessary to remove the solar
spectrum which contamination from a young moon.  To simulate most
accurately the types of kinematic data being acquired for high-redshift
galaxies, we summed the CCD rows over the observable spatial extent of
each galaxy to produce a 1-dimensional spectrum.  Several extraction
apertures of varying size were tested, and produced essentially
identical results except in the rare cases where bright H~II regions at
large radii generate much of the emission line luminosity.  In these
cases, care was taken to include the entire galaxy in the 1-D spectrum
for both the fixed-slit exposures and the drift scan exposures, even
though the larger extraction apertures produce 1-D spectra with lower
signal-to-noise.  We ultimately used the 1-D spectra with smaller
extraction windows for analysis.  Since the 1-D spectra are effectively
intensity-weighted global spectra, the high surface brightness regions
at small radii produce most of the kinematic signature.
 Our results for the small and large apertures yield comparable
kinematic measurements, but the smaller extraction window yields more
robust results due to the higher signal-to-noise.

In the analysis that follows, we convert all velocities to the
heliocentric frame of reference.  All velocities and velocity widths
discussed in this work refer to  projected velocities (in the plane of
the sky).  In order to compare observable quantities most directly,
no assumptions or corrections for galaxy inclination have been made.

\section{Analysis}

\subsection{Two-D [O~II] Emission Line Rotation Curves}

Courteau (1997) discusses the merits and shortcomings of several
methods for extracting rotation curves from spectral data.  Here we
adopt a straightforward approach, constructing a rotation curve by
fitting the centroids of the [O~II] emission features along the spatial
dimension.  The size of each spatial bin was varied to ensure
approximately equal number of counts in each bin.  At each spatial bin
we fit the [O~II] doublet using a Gaussian broadening function
convolved with the instrumental profile.  Calibration lamp exposures
provided a measurement of the instrumental spectral profile.  Based on
Gauss-Hermite polynomial fitting, the instrumental profile has very
little skewness ($|h|_3<0.02$) and tail weight ($|h|_4<0.04$), allowing
us to model it as a single Gaussian shape with $FWHM=1.57$ \AA\ (126
\kms\ at 3727 \AA).  Since the rotation curve  estimates require only
knowledge of the velocity centroid, a slight misreprentation of the
instrumental or broadening function does not significantly affect the
results.  Allowing the relative strengths of the [O~II] doublet to vary
over reasonable values ($0.5 < I_{3729}/I_{3727}<1.4$) produced the
same rotation curve to within the formal uncertainties.

The resulting rotation curves appear in Figure~2.  NGC~3044 is an
example of a classical spiral galaxy rotation curve that is
well-defined along many spatial resolution elements.  He~2-10 is an
example of a compact galaxy and shows little detectable rotation at
this spatial and spectral resolution.  Table~3 lists the (projected)
velocity of maximum rotation, $V_{max}(O~II)$, as measured from the
data in Figure~2.  This estimate, and its uncertainty reflects the mean
and standard deviation of the last $n$ points in the rotation profile,
where the number n ranges from 1 to 5.  As discussed in Courteau
(1997), the best estimate results from parameter fitting, but the
differences from our approach are not significant for
the goals of this paper.

\subsection{1-D [O~II] Emission Line Profiles}

The integrated 1-D spectra result from a convolution of the [O~II]
doublet profile with the line-of-sight Doppler velocity profile.  A
deconvolution analysis is required to reconstruct the intrinsic
velocity profile.  Since the velocity profile shapes range from the
classic double-horn to a unimodal shape, the deconvolution must
accommodate this large variety. We use maximum-penalized likelihood
techniques to determine a non-parametric estimate of the velocity
profile. The approach taken here is similar to that used in Saha~\&
Williams (1994), Merritt (1997), and Gebhardt~\etal\ (2000). As a first
step, we construct the un-broadened [O~II] doublet profile based on the
instrumental spectral profile which is well-characterized by a Gaussian
function with $\sigma=0.67$ \AA\ (53~\kms\ for [O~II]).  We begin the
deconvolution with a trial velocity profile composed of 54 {\kms}--wide
velocity bins. We then convolve that profile with the measured [O~II]
instrumental doublet profile, and compare to the galaxy spectrum. The
program varies the bin heights of the trial velocity profile to
converge upon the best convolved galaxy spectrum, assessed using the
$\chi^2$ statistic.  The addition of a penalty function to the $\chi^2$
ensures smoothness in the velocity profile. The integrated squared
second derivative represents the penalty function (as in Merritt 1997).
We choose a smoothness parameter that weights the penalty function
relative to the $\chi^2$. This choice is partly subjective, however, a
broad range of smoothing values exists over which there is little
change in the derived velocity profile. More rigorous techniques for
optimizing the smoothing parameter involve generalized cross validation
(Wahba 1990; Silverman 1986), but are not important for our needs.  The
ratio of the [O~II] doublet, $I_{3729}/I_{3727}$ was allowed to vary
from the low density limit of 1.4 to a high density limit of 0.5 in
order to achieve a best fit.  In most cases, ratios less than 1.2
(electron densities greater than $\sim100$ $cm^{-3}$) produced
unacceptable fits, consistent with observations that most \HII\ regions
exhibit line ratios indicating low electron densities (Osterbrock
1989).

Table~3 summarizes the derived velocity parameters, including the
widths at 20\% of the peak flux, $W_{20}(O~II)$, and systemic
velocities, $V_0([O~II])$ (defined as the midpoint between the 20\%
velocities).  Where the profile is Gaussian, $W_{20}(O~II)$ can be
compared to other width measurements such as the full width at half
maximum (FWHM) and velocity dispersion, $\sigma$, using

\begin{equation}
\sigma={{FWHM}\over{2.35}} = {{W_{20}}\over{3.62}}
\end{equation}

\noindent Figure~3 displays the deconvolved [O~II] velocity profiles
for each galaxy.  Profiles from the fixed slit spectra appear in dotted
lines,
and the drift scan spectra appear in long dashed lines.  Although the lower
signal-to-noise of the drift scan spectra results in slight differences
between the line profiles, the correspondence between the fixed slit
and drift scan spectra is generally excellent.   Exceptions are
NGC~1741 where the fixed slit seems to have been offset from the
dominant emission line region.  More anomalies and exceptions are
discussed in \S\ 4 below.

We estimate uncertainties on the emission line velocity centers and
widths using a Monte-Carlo code which adds Gaussian noise to the
best-fit velocity profile convolved with the instrumental profile
and re-fits the profile.  We generate 100 such
realizations to characterize the distribution of probable values and
estimate confidence limits.  Table~3 includes 1 $\sigma$ uncertainties
on the line centers and widths based on the Monte Carlo simulations.

\subsection{1-D Stellar Absorption Line Profiles}

As another probe of galaxy kinematics, we use the stellar Ca H\&K and
Balmer absorption lines to measure the {\it stellar} velocity profile
of each galaxy.  The procedure uses the same maximum-penalized
likelihood approach as the emission-line analysis.  The only difference
is that we must use stellar templates instead of an instrumental
profile since the intrinsic widths of the absorption features are
broader than the instrumental profile.  Furthermore, galaxies consist
of a range of stellar populations that have dramatically different
spectral features and shapes, making absorption line work more
complicated than emission line studies.  To obtain reliable kinematic
measurements, one must separate the effects of a mixed stellar
population from the effects of broadening due to bulk stellar motions
within the galaxy.  This separation is particularly difficult in the Ca
H\&K region, and is why these features have not been widely used for
kinematic measurements.

Since the targets encompass a broad range of morphological types, we
must adequately sample the full range of stellar types in the host
galaxies.  The spectra of 12 Galactic A main--sequence through K giant
stars served as templates (see Table~2).  The maximum-likelihood
technique simultaneously determines the best velocity profile and the
relative contribution each stellar type.  As before, we use the
integrated squared second derivatives as the penalty function.  The
analysis software adjusts the velocity profile bin heights and the
relative weight to minimize the chi-squared fit to the data.  Due to
the prominent, but narrow, emission lines in some galaxies in the Ca H
and H$\epsilon$ vicinity, the emission line regions are excised from
the spectrum before the fit. Figure~4 shows the 3850 \AA\ -- 4040
\AA\ region of each galaxy spectrum with prominent absorption
features.  Three strong emission lines are present in this range
including the blended [Ne III] $\lambda$3868/He I $\lambda$3867, He I
$\lambda$3889 bended with H8, and the H$\epsilon$ blended with [Ne~III]
$\lambda$3967.  Figure 4 shows the raw spectrum, along with with the
best fit synthesis of template stars convolved with the estimated
velocity profile overplotted in solid lines.  Dashed lines show excised
regions contaminated by emission features.   Some galaxies with strong
emission lines such as NGC~4214 require that extensive regions be
ignored in the fit.  This leads to larger uncertainties on the derived
kinematic parameters.  The choice of excised regions often depends on
the particulars of the spectrum.  In general we can fix the excised
regions: if emission lines are present, we exclude them (3864 \AA\ --
3872 \AA, 3885 \AA\ -- 3892 \AA, 3965 \AA\ -- 3973 \AA).

One spectral region was consistently difficult to fit with any
combination of template star spectra.  The region from 3900 \AA\ --
3940 \AA\ showed significant residuals in the best fit synthesis
spectra.  Two different problems occur.  The galaxies He~2-10,
NGC~2276, NGC~2798, NGC~4214, and NGC~5248 show strong positive
residuals near 3910 \AA\, resembling emission features.  The origin of
these weak emission features in some galaxies is not clear.  Such
stellar features do not appear in any of our template spectra, nor are
there any known nebular emission lines in this region.  One possible
source of extraneous lines in this portion of the spectrum is
chromospheric emission from giants and binary systems (Stencel 1975;
Strassmeier \etal\ 1993).  However, these features are usually weak and
would probably not appear at a significant level in the integrated
spectrum of an entire galaxy.  Another possibility is CN molecular
bands in giant stars, but this effect should produce a larger signature
in late type galaxies, contrary to the observed trend.  We have not
been able to find a satisfactory explanation for the origin of these
features. We decided on a case-by case basis whether to exclude this
region.  When no emission lines are present, we use the full spectral
range (e.g., NGC~4527, NGC~4666).  Another difficulty involves
fitting template spectra to galaxies like NGC~2798, NGC~3044, NGC~4218,
and NGC~4818, where the Ca K line is narrower and deeper than any
combination of template stars can match.    Either of these
difficulties are plausibly due to metallicity mismatches between the
template stars and the galaxies studied.  The stellar templates in
Table~2 do not include metal-poor stars which are probably common in
some of the smaller star-forming galaxies like NGC~4214.

Figure~3 plots the resulting absorption line velocity profiles for each
galaxy.  We record in Table~3 the velocity width at 20\% of the maximum
profile depth, and the systemic velocity defined here as the midpoint
between the 20\% velocities.  We estimate uncertainties on the
velocities and widths using the same Monte-Carlo technique described in
\S\ 3.2.

Sixteen of the 22 targets had sufficient signal-to-noise and measurable
absorption line features for analysis.  Five of the six galaxies
without detectable absorption features (NGC~1741, Mrk~1089, NGC~5253,
UM~439, and UM~462) exhibit strong nebular emission which dominates the
spectrum.  The low continuum signal-to-noise combined with a strong
background solar spectrum from the crescent moon precluded a meaningful
measurement for NGC~925.

\subsection{21-cm Neutral Hydrogen Spectra}

We compiled from the literature the single-dish 21-cm neutral hydrogen
spectrum for each galaxy.  When more than one profile was available, we
selected the one with the best combination of signal-to-noise and
velocity resolution, (typically $\sim11$ \kms).  Most single-dish
observations are based on data with the 92 m NRAO Greenbank
radio-telescope, which has a FWHP beamsize of 10\arcmin, large enough
to adequately cover the angular extent of galaxies in our sample.  In
order to compare the 21-cm profiles with the newer optical data, we
digitized the original published spectra to generate a spectrum in
electronic form.\footnote{In most cases, the 21-cm spectra date from the
early 1980's, and the data are not recoverable on electronic media from
the original authors.}  From these digitized spectra we measured the
full-widths at 20\% of the peak intensity, $W_{20}(H~I)$, a standard
parameter commonly tabulated in 21-cm studies.  We also computed the
systemic velocity, $V_0(H~I)$, defined here as the midpoint between the
20\% velocities.  Table~3 lists the original publications for each
21-cm spectrum, along with 21-cm systemic velocities and
$W_{20}(H~I)$.  Our re-measurements of the systemic velocities and
velocity widths from the digitized data are in good agreement with the
published values.  We estimate that the digitization process introduced
errors of $<5$ \kms\ in velocity and $<$5\% in amplitude compared to
the original data.  The resulting \HI\ spectra appear in Figure~3.
There are no published \HI\ spectra  for NGC~4818 but only marginal
detections.

To complement the 1-D 21-cm profiles, we searched the literature for
spatially resolved rotation curve measurements from aperture synthesis
observations.  Table~3 lists the maximum \HI\ rotational velocity,
$V_{max}(HI)$, for all galaxies with suitable data, as found in the
original references.  If more than one aperture synthesis measurement
was available, we adopted the most sensitive study, usually from recent
Very Large Array (VLA) programs.

\section{Results}

\subsection{Comparison of Velocity Widths
Measured from \HI\, [O~II], and Stellar Absorption Features}

One of the goals of this program was to search for systematic
differences between morphological types or pathological systems of any
type which exhibit markedly different kinematic signatures in one or
more of the measured tracers.  Figure~3 shows graphically the
one-dimensional velocity profiles measured from 21-cm, [O~II], and
stellar absorption line spectroscopy.  The 21-cm profiles appear in
solid lines, the fixed slit [O~II] profiles in dotted lines, the drift
scan [O~II] spectra in long dashes, and the absorption spectra in short
dashes.  A cursory comparison by eye reveals that the overall systemic
velocities and velocity widths measured from the different tracers are
generally in good agreement.  We discuss exceptional systems in more
detail below.  Henize 2-10 and Mrk 33 are example of low-mass systems
with Gaussian velocity profiles where the three dynamical indicators
are in excellent agreement.  NGC 2798 and NGC 3044 are examples of
larger galaxies where all three measurements yield comparable results,
even though the 21-cm and [O~II] profiles are double-peaked and the
absorption profile from the stellar component is single-peaked.

Figure~5 compares the derived velocity widths from each of the three
kinematic tracers.  In the upper right panel we show the the [O~II]
full width at 20\% max, $W_{20}([O~II])$, versus $W_{20}(H~I)$.  Filled
symbols distinguish the fixed-position spectra from drift scan spectra
(open symbols).  Dashed lines denote 20\% deviations from the 1:1
correspondence (solid line).  NGC~4818 and Mrk~1089 could not be
plotted since they do not have published neutral hydrogen spectra.
Figure~5 shows a strong correlation between $W_{20}([O~II])$ versus
$W_{20}(H~I)$, most galaxies falling within 20\% of the 1:1 relation.
The drift scans yield results consistent with the fixed slit exposures,
within the uncertainties.  NGC~4449 and  NGC~4666 stand out as having
broad \HI\ widths compared to the [O~II] profile both here and in
Figure~3.  Inspection of their spectra in Figure~3 reveals that
NGC~4449 exhibits a small kinematic width in [O~II] perhaps because the
\HII\ regions are concentrated near the center of an extended neutral
gas distribution and trace only a small fraction of the gravitational
potential.  NGC~4666 is a starburst galaxy with low-level [O~II]
emission across the entire \HI\ velocity width, and a strong emission
line concentration on one velocity wing.  This emission is confined to
one portion of the galaxy, and does not trace the full gravitational
potential, giving the false signature of a less massive galaxy.  If the
spectra had been of lower S/N, the weak [O~II] emission across the
entire 300 \kms\ velocity width seen in \HI\ would have gone
unrecognized.  NGC~2966  exhibits slightly smaller \HI\ widths compared
to the [O~II] profile and stellar profile.  Unfortunately, there is
only one published \HI\ profile for this galaxy, so we are unable to
explore the possibility that the neutral hydrogen width is somehow
underestimated.

The upper left panel of Figure~5 compares the \HI\ line widths with the
absorption line widths.  There is again a good correlation
between the two indicators.  Within the often substantial uncertainties
on the absorption line widths, all the data are consistent with the 1:1
correspondence except the irregular galaxy NGC~4214.  The Seyfert
galaxy NGC~1068 is also discrepant, but since we are unable to
spatially separate the high-velocity nuclear regions probed by stellar
absorption from the rest of the galaxy, agreement is not expected.  We
do not consider NGC~1068 further, except as an example of an AGN where
the nuclear kinematics are distinct from the overall gravitational
potential of the galaxy.

The lower right panel of Figure~5 compares the [O~II] emission versus
stellar absorption velocity widths. Once again NGC~4666 and NGC~4449
and NGC~4214 stand out as deviant points for the reasons discussed
above.  The rest of the data show good agreement between the two
kinematic indicators.  

As an alternative measure of the velocity width, we computed the
difference between velocities where the integrated area under the
profile on either side reaches 10\% of the total area (see Courteau
1997 for discussion of this and other methods).  This approach is more
robust when measuring objects like NGC~4666 because it is not as
sensitive to instrumental resolution or asymmetric profiles.  However,
this approach is prone to larger uncertainties for noisy data.  The
[O~II] width of NGC~4666 changes from 170 \kms\ to 392 \kms\ when
measured in this manner, bringing the [O~II] width into accord with the
other dynamical indicators.  In general, the measured linewidths of all
target galaxies become 15\% to 25\% smaller using this approach.  The
correlation between the various dynamical indicators explored here
remains consistent with the above discussion.

\subsection{Comparison of Rotation Curves
Measured from \HI\, [O~II], and Stellar Absorption}

When spatially-resolved velocity data are available, rotation curves
provide more information about a galaxy's kinematics than an integrated
profile alone.  In each of the target galaxies we measured [O~II]
rotation curves and the maximum rotational velocity,
$V_{max}([O~II])$.  Table~3 lists these values.  Figure~6 shows a
comparison of the projected $V_{max}([O~II])$ with the 21-cm rotation
curve maximum, $V_{max}(H~I)$ where such data were available in the
literature.  We also compare the $V_{max}$ measurements to estimates of
$V_{max}$ based on $W_{20}(H~I)$ from the prescription of Tully \&
Fouqu\'e (1985),

\begin{equation}
W_R^2 = W_{20}^2 + W_t^2 - 2W_{20}W_t[1-e^{-(W_{20}/W_c)^2}] - 2W_t^2
e^{-(W_{20}/W_c)^2}.
\end{equation}

\noindent Here, $W_R$ is the rotation full amplitude which is 2$\times
V_{max}$.  $W_t=38$ \kms\ is the width due to turbulent motions and
$W_c=120$ \kms\ is the transition point between galaxies having
Gaussian and those having double-horned \HI\ profiles.  We see in
Figure~6 that the maximum points on the rotation curves measured either
from neutral hydrogen aperture synthesis data or [O~II] are
systematically less than those predicted by simply taking half of
$W_{20}(H~I)$.  There is better agreement with the analytic expression
of Tully \& Fouqu\'e which takes into account the effects of turbulence
that become more important in the smallest galaxies.  The most
discrepant points are the compact low-mass galaxies like He~2-10,
Mrk~33, and NGC~925 where measuring a rotation curve becomes difficult
due to limited spatial resolution, or because the intrinsic line width
due to turbulent motions becomes comparable to the (projected)
rotational velocity.  The [O~II] rotation curve may easily be
underestimated due to poor sensitivity or spatial resolution,
especially at large distances.

\subsection{Comparison of Systemic Velocities
Measured from \HI\, [O~II], and Stellar Absorption}

As a measure of the systemic velocity, $V_0$, for each system, we
record in Table~3 the velocity midway between the 20\% intensity points
used to define $W_{20}$.  This formulation of the systemic velocity is
a well-defined quantity in the case of massive galaxies with two-horned
velocity profiles.  It is also less sensitive to asymmetries in the
line profile in the case of low-mass galaxies with Gaussian profiles
where most of the nebular emission may come from a single
\HII\ region.  Figure~7 illustrates the differences between systemic
velocities measured with each technique.  We show, as a function of the
(projected) neutral hydrogen width, the difference between the
$V_0(H~I) - V_0(Abs)$ (upper left panel), $V_0(H~I) - V_0([O~II])$
(upper right panel), and $V_0([O~II]) - V_0(Abs)$ (lower right panel).
Open symbols distinguish drift scan exposures from fixed-slit
integrations (solid symbols).  Overall, the three methods yield similar
velocities for most galaxies to within the uncertainties on the fits.
In nearly all cases the drift scans show better agreement than the
fixed slits.  In the upper left panel, there is good agreement between
the systemic velocities measured with \HI\ and stellar features.

Turning to the upper right panel of Figure~7, a few galaxies show real
differences discrepancies as large as 50 \kms\ between the \HI\ and
[O~II] velocities.  In the case of NGC~4527, $V_0([O~II])$ exceeds
$V_0(H~I)$ by 70 \kms.  Since we do not have a drift scan of this
object, we suspect that the fixed slit position missed much of the
emission line gas.   The line width derived from the [O~II] spectrum is
narrower ($W_{20}=409$ \kms) than the Lehnert \& Heckman [N~II] width
of $W_{20}=500$ \kms.  Their rotation curve amplitude of 339 \kms\ is
significantly larger than our 260 \kms, consistent with the probability
that our choice of slit placement did not cover the entire
emission-line extent of the galaxy.  Other galaxies show only marginal
evidence for systematic offsets between the neutral and ionized gas at
the level of $<10$ \kms.

\subsection{Comments on Individual Objects}

{\it Henize 2-10:\ } Henize~2-10 is a blue compact galaxy with a solid
body rotation curve typical of dwarf and compact objects (Kobulnicky
\etal\ 1995).  Figure~3 shows that the \HI, and [O~II] features
indicate a similar width of $W_{20}=156\pm15$ \kms\ and
$W_{20}=158\pm10$ \kms.  The measured width of stellar features is
slightly greater but consistent within the uncertainties
$W_{20}=205^{+132}_{-70}$ \kms.  The systemic velocity as measured by
the \HI\ spectrum appears  consistent with the optical data.  Given
the strong star formation and high optical depth, it is likely that
only the nebular emission on the near side of the galaxy is seen in the
optical spectra.

{\it Markarian 33:\ } Another blue compact galaxy, Mrk~33 shows an
\HI\ line width ($W_{20}=208\pm25$ \kms) which is 30 \kms\ larger than,
but consistent with, the nebular emission ($W_{20}=168\pm5$
\kms) and the stellar line width ($W_{20}=187^{+58}_{-48}$ \kms).  The
emission line gas is evidently confined to only a fraction of the
rotation curve traced by the neutral hydrogen and stellar components as
suggested by the 1-D profiles in Figure~3.  The [O~II] rotation curve
has a very small measurable amplitude compared to the intrinsic line
width measured in the optical and neutral gas, causing Mrk~33 to stand
out in Figure~6.  This small rotation amplitude is most probably a
result of limited spatial resolution and the relatively larger distance
of Mrk~33. 

{\it NGC 925:\ } This is a moderately inclined late-type spiral for
which we can only measure [O~II] in the optical spectrum.  The stellar
absorption features are weak, and at the low redshift of 550 \kms, its
Ca and Balmer features are strongly contaminated by the solar spectrum
produced by a nearby 15\% illuminated moon.  The emission line profile
is narrow ($W_{20}=176\pm7$ \kms) compared to the \HI\ profile
($W_{20}=222\pm6$ \kms).  The [O~II] and \HI\ systemic velocities
agree well near $V_0=555\pm3$ \kms.

{\it NGC~1068:\ } As the only AGN (Seyfert 2 class) in our sample,
NGC~1068 shows a broad nuclear emission component which produces the
unusual rotation curve in Figure~2.  The drift scan exposure yields a
lower velocity width than the fixed slit which is dominated by the
nuclear kinematics, so we consider only the former.  On the basis of
the integrated 1-D  optical spectrum, it is not possible to deduce a
velocity width that accurately reflects the rotational velocity of the
outer disk.  The measured emission line width is $W_{20}([O~II])=1129\pm200$
\kms\ compared to the much lower stellar ($W_{20}(Abs)=564^{+38}_{-59}$
\kms) and \HI\ kinematics ($W_{20}(\HI)=298\pm30$ \kms).  Because of the
unique nature of the this object, we do not consider it further for
analysis, except to note this as an instance where the nebular lines
would clearly lead to an erroneous estimate of the dynamical mass of
the system.

{\it NGC~1741/Mrk~1089:\ } These galaxies are members of a Hickson
Compact Group (HCG 31) components a and c respectively.  Both objects
exhibit strong emission lines observed simultaneously with a single
slit placement.  A comparison of the drift scan exposure with the fixed
slit shows that, in the case of NCG~1741, a small displacement of the
slit from the dominant emission line region of the galaxy can yield a
discrepant systemic velocity.  The available \HI\ profiles encompass
both galaxies within the large beam of typical radio telescopes.  VLA
aperture synthesis maps (Williams \etal\ 1991) reveal a complex
\HI\ distribution indicating that these galaxies are interacting and
are possibly connected by a common pool of neutral hydrogen.  Most of
the neutral gas is affiliated with NGC~1741, with secondary peaks in
Mrk~1089 and the other two members of the group.  This configuration
explains why the \HI\ spectrum in Figure~3 reflects more closely the
[O~II] kinematics of NGC~1741 than Mrk~1089 (not plotted).  Comparisons
between the nebular and \HI\ kinematics of Mrk~1089 are thus
inappropriate until a separate \HI\ profile of Mrk~1089 is available.
It was not possible to discern stellar absorption features in either of
these objects.

{\it NGC~2146:\ } This early type spiral has the largest velocity
amplitude in our sample.  Figure~5 shows that the \HI\ width is in good
agreement the nebular and stellar tracers.  The systemic velocities
measured by each tracer are in good agreement for the drift scan
measurements, as are the maximum rotation speeds measured from the
\HI\ and [O~II] rotation curves.  Systemic velocities measured from the
fixed slit only yield [O~II] centroids which are 60 \kms\ smaller than
the stellar and 21-cm systemic velocities.  This result underscores
potential difficulties with using single-position slits rather than
drift scans to measure recessional velocities of spatially-resolved
galaxies.

{\it NGC 2276:\ }  A moderate-inclination late-type spiral, NGC~2276
shows good agreement between all three kinematic width tracers in
Figure~5.  In Figure~7, and from the profiles plotted in Figure~3, we
note a 20 \kms\ offset between the neutral hydrogen profile and the
optical [O~II] and stellar profiles.  The $H\alpha$ image in Young
\etal\ (1996) reveals a potential reason for this offset.  In NGC~2276,
\HII\ regions and stellar light preferentially occupy the western half
of the galaxy, so that the (intensity weighted) 1-D [O~II] and stellar
absorption profiles are biased away from the systemic velocity.  This
object serves to illustrate the magnitude of scatter introduced by
inhomogeneous distributions of stellar and gaseous components.

{\it NGC 2798:\ } \HI\ and stellar velocity widths in this
low-inclination Sa galaxy are consistent near $310\pm30$ \kms, but the
optical emission line velocities are somewhat broader at 387$\pm21$
\kms. The measured systemic velocities agree well in Figure~7 within
the uncertainties.  These substantial errors are due to the relatively
low-S/N.  The H$\alpha$ profile of the nuclear regions which dominates
the spectrum is (Lehnert \& Heckman 1995)\footnote{Lehnert \& Heckman
report [N~II] profiles in terms of the full width at half maximum
(FWHM) and the rotation curve parameters in terms of the rotation full
amplitude, $A_{rot}$.  For comparison, we convert these parameters to
20\% widths and rotation curve maxima using $W_{20}=1.54\times{FWHM}$
and $V_{max}=0.5\times A_{rot}$.} $W_{20}(H\alpha)=362$ \kms,
consistent with our measurement of $W_{20}([O~II])=387\pm21$ \kms.

{\it NGC 2966:\ } This low-inclination spiral exhibits the classic
double-horned velocity profile in both \HI\ and in [O~II].   Both the
stellar absorption and [O~II] profiles are 20-40\% broader than the
\HI\ profile $W_{20}(H~I)=258\pm25$ \kms.  Unfortunately, there is only
one published 21-cm spectrum of NGC~2966, so we are unable to verify
the \HI\ parameters.   The drift scan [O~II] data also yields an
unusually broad profile compared to the fixed slit.    Lehnert \&
Heckman (1995) present an H$\alpha$ image which reveals a nuclear
emission region with additional \HII\ regions in the spiral arms at
large radius.  Their H$\alpha$ longslit spectroscopy yields a FWHM of
257 \kms\ for their largest measurement, which would correspond to
$W_{20}=395$ \kms, marginally consistent with our
$W_{20}([O~II])=362\pm15$ \kms.  Their maximum projected rotational
velocity, is 116 \kms\ which is smaller than our 150$\pm$10 \kms.
Given the agreement between our optical results and those of Lehnert \&
Heckman, we are suspicious of the low \HI\ width and would urge a new
single-dish measurement of this galaxy.

{\it NGC 3044:\ } As one of the largest and brightest galaxies in our
sample, NGC~3044 has a well defined rotation curve and velocity profile
in every tracer.  The profiles in Figure~3, and the kinematic summaries
in Figures~5 and 7 shows that the shows that the \HI\ and [O~II]
velocity widths are in good agreement near $W_{20}=360\pm10$ \kms.
However, the velocity profile measured from the absorption lines is
substantially more narrow, $W_{20}(Abs)=279^{+79}_{-66}$ \kms.  This
appears to be an instance where the light-weighted stellar spectrum
samples a smaller portion of the rotation curve than neutral hydrogen
and ionized gas at larger radius.  
The systemic velocities are in excellent agreement.

{\it NGC 3593:\ } In this Sa type spiral, linewidths of the \HI\ and
stellar absorption agree well in NGC~3593, $W_{20}\sim300$ \kms, while
the optical emission lines are much broader, $W_{20}([O~II])=411\pm25$
\kms.  This spectrum has low S/N and fairly large uncertainties.  Our
optically-measured line widths of $W_{20}=300-400$ \kms\ are much
larger than the 230 \kms\ widths implied by the [N~II] spectroscopy of
Lehnert \& Heckman (1995).  However, a direct comparison of the
velocity widths is not possible since  Lehnert \& Heckman subdivide
their spectra in to several different radial bins whereas we consider
only the entire spatially integrated profile.  Despite this possible
difference line widths, the maximum [O~II] circular velocity of
$V_{max}=103\pm10$ \kms\ is roughly consistent with the [N~II] circular
velocity of 118 \kms\ reported by Lehnert \& Heckman.  The systemic
velocities measured from all three indicators are in excellent
agreement.

{\it NGC 4214:\ }  This Magellanic irregular has one of the narrowest
velocity profiles in our sample, $<100$ \kms, at the limit of our
resolution.  The \HI\ width is only 82 \kms, comparable to the optical
results which suggest  $W_{20}([O~II])\sim60$ \kms.  Multiple results
from the literature confirm this low neutral hydrogen width.  The
systemic velocities in Figure~7 show good agreement.  Curiously, the
stellar velocity profile is much broader, $W_{20}(Abs)=132^{+55}_{-27}$
\kms, making NGC~4214 one of the most outstanding points in Figure~5.

{\it NGC 4218:\ } The velocity centroids and widths of all measurements
are in good agreement for this moderately inclined Sa type spiral,
ranging between $W_{20}=160-190$ \kms.  The \HI\ profile is centered
centered 20-30 \kms\ lower than the optical results.  The velocity
widths of the [O~II] emission lines are broad, $W_{20}([O~II])=195\pm6$
\kms\ compared to the absorption 21-cm width of $163\pm20$ \kms.
Aperture synthesis measurements of the 21-cm profile yield a
surprisingly narrow $W_{20}=138$ \kms\ (Verheijen 1997) while single
dish profiles range between 157 \kms\ and 201 \kms, consistent with the
optical results.

{\it NGC 4449:\ } Similar in many ways to NGC~4214, this irregular
galaxy has an unusually large \HI\ halo extending to 14 times the
optical radius (Bajaja \etal\ 1994).  The \HI\ and stellar absorption
profiles agree well, near  $W_{20}\simeq190$ \kms,
 while the [O~II] fixed and drift exposures are considerably more
narrow, $W_{20}=90\pm21$ \kms.  Since the signal-to-noise is high in
both the optical and 21-cm spectra, this difference s a robust result.
NGC~4449 is an example of an object where the global 1-D
(intensity-weighted) emission line profile traces only a portion of the
rotation curve.  Rotational kinematics based only on an [O~II] profile
would underestimate the rotational velocity by 50\%.  NGC~4449 is the
most significant outlier in Figure~5.  Figure~7 shows that the systemic
velocities measured from each method are in good agreement.

{\it NGC 4527:\ } One of the most massive galaxies in our sample,
NGC~4527 shows good agreement between the absorption line
and \HI\ kinematics, $W_{20}=370$ \kms.   Lehnert \& Heckman report an
[N~II] width of $W_{20}=500$ \kms\ in the nuclear region.  Their
rotation curve amplitude of 339 \kms\ is significantly larger than our
232$\pm$13 \kms, suggesting that our choice of fixed slit placement did
not cover the entire emission-line extent of the galaxy.  The [O~II]
systemic velocity is larger by 60 \kms\ but carries large uncertainties
due to the presence of low-level emission/noise at extreme velocities
(see Figure~3).

{\it NGC 4666:\ } Another large, edge-on galaxy, NGC~4666 exhibits good
agreement between the velocity widths measured in \HI\ and stellar
features near $W_{20}=400$ \kms.  The double-horned \HI\ profile
contrasts with the single-peaked stellar absorption profile as shown in
Figure~3.  The [O~II] profile shows one dominant emission peak
and superimposed on low-level emission.  The dominant emission peak centered  near 1700
\kms\ gives the misleading appearance of a narrow velocity profile with
$W_{20}([O~II])=170\pm13$ \kms.  NGC~4666 serves as a warning that in a
low S/N spectrum of a typical high-redshift galaxy, 
this object could appear to have a
small kinematic width.  Figure~5 shows that the measured [O~II] profile
is less than half as large as the \HI\ profile.  Unfortunately we do
not have a drift scan observation of this target to verify that an
integrated spectrum would produce the same result.  This unusual
emission line profile leads to an artificially aberrant emission line
centroid, displaced toward higher velocities.  

{\it NGC 4818:\ } There is no published \HI\ spectrum for this galaxy,
and only recently is a detection reported by Theureau \etal\ (1998).
The absorption line profile is slightly more narrow than the
emission line profile ($231\pm80$ \kms\ versus $276\pm45$ \kms) in this
early type spiral.  

{\it NGC 5248:\ } Here the \HI, [O~II] and stellar profiles give very
similar results, $W_{20}=293-301$ \kms, for this Sbc galaxy.  Likewise
the systemic velocities agree well.  Like NGC~3044, the stellar profile
is single peaked, while the \HI\ and [O~II] profiles are
double-horned.

{\it NGC 5253:\ }  NGC~5253 is an amorphous galaxy with a strong
central starburst and nebular emission. No stellar features are
measurable.  The \HI\ and [O~II] profiles are Gaussian and agree well
in their velocity widths.
Lehnert \& Heckman (1995) find a maximum rotational amplitude of 3
\kms\ amidst the distorted and irregular velocity field.  Martin \&
Kennicutt (1995) show ordered velocity variations of 25 \kms\ along the
{\it minor} axis, consistent with the unusual minor-axis
\HI\ kinematics seen in VLA maps (Kobulnicky \& Skillman 1995).

{\it UM 462:\ } A blue compact dwarf galaxy, the spectrum of UM~439 is
dominated by emission lines and no stellar features are measurable.
The [O~II] rotation curve is consistent with solid body rotation and is
measurable over 25 pixels (1.4 kpc).  The velocity widths in
\HI\ and [O~II] are in good agreement.

{\it UM 439 :\ } As with UM~462, the velocity profiles are Gaussian
and well-matched in both \HI\ and  [O~II] with $W_{20}\simeq100$ \kms.
No stellar features are measurable.

\section{Discussion and Conclusions}

We have examined the rotation curves, systemic velocities, and 1-D
velocity widths for a wide variety of galaxy types and inclinations to
search for systematic deviations between measurements made with 21-cm,
nebular [O~II], and stellar absorption spectra.   The three methods
provide kinematic line widths which agree to within 20\% for most
(19/22) objects, and within 50\% in all galaxies studied.  NGC~4449, an
irregular galaxy with an extremely extended neutral hydrogen halo, is
the most discordant: the [O~II] line width under predicts the 21-cm
line width by a factor of 1.6.  NGC~4666 also shows broad extended
nebular emission and a strong narrow [O~II] peak which might be
interpreted as a signature of a small dynamical mass had the
signal-to-noise been lower.  NGC~4666 serves as a warning that in a low
S/N spectrum of a typical high-redshift galaxy, this object could
appear to have a small kinematic width.

In general, however, the stellar kinematics as traced by the absorption
features yield line widths which are nearly as reliable as the [O~II]
doublet, even in many late-type galaxies with strong nebular emission
lines.  Due to the patchy distribution of \HII\ regions within some
galaxies, the stellar features yield more reliable estimates of the
systemic velocities (as measured by \HI\ 21-cm spectra) than the [O~II]
data.  The spatially-integrated drift exposures yield better estimates
of the velocity widths and systemic velocities.  The intrinsic
dispersion in the difference between systemic velocities measured from
the gas and stars appears to be less than 10 \kms, comparable to our
measurement uncertainty, in most systems.

Although our sample does not include early-type galaxies, we expect
that the Balmer and Ca H\&K region will prove suitable for
absorption-line kinematic analysis in those systems as well.  For the
galaxies that show no visible emission lines, we were able to use the
complete spectrum without needing to excise any spectral region. In
these cases, the agreement with the \HI\ profiles was excellent.  Since
early-type galaxies generally do not show emission features, the full
optical spectrum may be used in a kinematic analysis.  However, we
advise caution when emission lines are present.   Spectral regions
contaminated by emission lines must be excised before measuring the
absorption width. Furthermore, in some galaxies, the region from 3900
\AA\ -- 3940 \AA\ is affected by a series of weak unidentified lines
which can cause difficulty at discussed in \S 3 above.  This region
should be excised if adequate template stars are not available.

One goal of this program was to determine whether galaxy masses derived
from [O~II] emission and stellar absorption features are intrinsically
noisier or biased compared to more traditional dynamical indicators. In
order to estimate the intrinsic scatter between \HI, [O~II], and
absorption-feature velocity widths in Figure~5, we compare the observed
RMS to the expected RMS from the formal uncertainties computed by Monte
Carlo methods.  The measured RMS between the \HI\ width and [O~II]
emission is 47~\kms, and for the absorption width it is 54~\kms. Given
the uncertainties, the expected RMS is 40 and 33~\kms, respectively.
Thus, the data require an additional intrinsic scatter beyond the
experimental measurement errors.   This additional scatter is
18~\kms\ added in quadrature to the [O~II] width uncertainties, and
20~\kms\ for the absorption widths.  Since a typical galaxy in our
sample has a velocity width greater than 200~\kms,  either kinematic
tracer provides a suitable measure of the width, better than 10\% in
velocity or 20\% in the mass (given that mass, $M\propto\Delta V^2$).
Surprisingly, this result is true using absorption line kinematics in
some of the very late-type galaxies where the mix of stellar population
causes large spatial variations in the spectrum.  For low mass galaxies
with narrow ($<150$ \kms) velocity widths, the measurement
uncertainties can approach 50\% in velocity and, thus, factors of 2 in
mass.  While these kinematic indicators should still be useful for
measuring the gravitational potential of low-mass galaxies, the
relative uncertainties will be significant.

For the most massive systems, however, uncertainties due to the
measured intrinsic scatter are smaller than, or comparable to, the
dispersion observed in local Tully-Fisher and Fundamental Plane
relations.  For Tully-Fisher, the scatter has been estimated to be 0.2
magnitudes (Pierce~\& Tully 1992; Tully \& Pierce 1999) to 0.25-0.4
magnitudes (Courteau 1997).  This translates into a 12.5\% to 20\%
error in distance or a 10\% -- 16\% error in velocity width (assuming
typical slopes, $a$, of 6--7 in the Tully-Fisher relation; $M=a [log V_c
-2.5]+b$ ).  Thus, the 18~\kms\ uncertainty, which may be introduced by
adopting [O~II] emission line widths or stellar absorption line widths,
only begins to add to the distance scatter for galaxies with W$_{20}$
below 200~\kms. Given that the Tully-Fisher galaxies are generally the
most massive since they are the easiest to observe and have
well-defined rotation curves (as used in Vogt~\etal\ 1997), this
additional scatter will have at most a small effect for distant
Tully-Fisher analysis.

For the Fundamental Plane, the scatter is around 8\% (J\o
rgensen~\etal\ 1998) in logarithmic effective radius, implying a
scatter of 0.25 magnitudes in surface brightness or a 15\% error in the
velocity dispersion. Since most of our additional uncertainties are
below 10\% for the absorption line widths, it will have little effect
on either the Fundamental Plane scatter or in the estimate of the
difference in surface brightness. Since both [O~II] emission and
stellar Balmer/Ca H\&K absorption linewidths correlate well with \HI\
21-cm linewidths we anticipate the increased use of these kinematic
indicators to derive galaxy masses in the distant universe.

\acknowledgments 

We are grateful for helpful conversations and suggestions from Matt
Bershady, Sandy Faber, Luc Simard, Drew Phillips, St\'ephane Courteau,
and the referee, Chris Pritchet.  We thank Liese van Zee for providing
the 21-cm spectra of UM~439 and UM 462, D.~J. Pisano for the 21-cm
spectra of NGC~925, Siow-Wang Lee for 21~cm spectra of NGC~3044 in
electronic form, and Benjamin Weiner for the image of NGC~1068.  
We thank Evan Skillman and David E. Hogg for their
help in tracking down older \HI\ spectra which were surprisingly
difficult to locate by electronic means.  Greame Smith and Jean Brodie
offered helpful advice on the nature of the troublesome stellar
features in the 3900 \AA\ -- 3930 \AA\ region.  H.~A.~K and K.~G. are
grateful for support from Hubble Fellowship grants \#HF-01094.01-97A
and \#HF-01090.01-97A awarded by the Space Telescope Science Institute
which is operated by the Association of Universities for Research in
Astronomy, Inc. for NASA under contract NAS 5-26555.

\clearpage

\begin{figure} 
\figcaption[fig1] {Optical images of the
21 target galaxies, along with the two-dimensional spectra.
The scale bar corresponds to 1\arcmin\ along the spatial direction.
The width of the image is  30 pixels (1050\kms) in the spectral 
direction with longer wavelengths to the right.
The dashed line show the extent of the drift scans.}
{ \label{multiplot} } \end{figure}

\clearpage

\clearpage

\begin{figure} 
\figcaption[fig2] { [O~II] $\lambda$3727,3729 rotation curves
for the galaxy sample.  The y axis shows the relative
velocity and the x axis is in arcseconds from the center of the galaxy.
The data points and error bars represent bins with equal flux.  
The dashed line show the maximum [O~II] rotational amplitude, $V_{max}$.}
{ \label{plotrc} } \end{figure}
\clearpage
 
\clearpage

\begin{figure} \centerline{\psfig{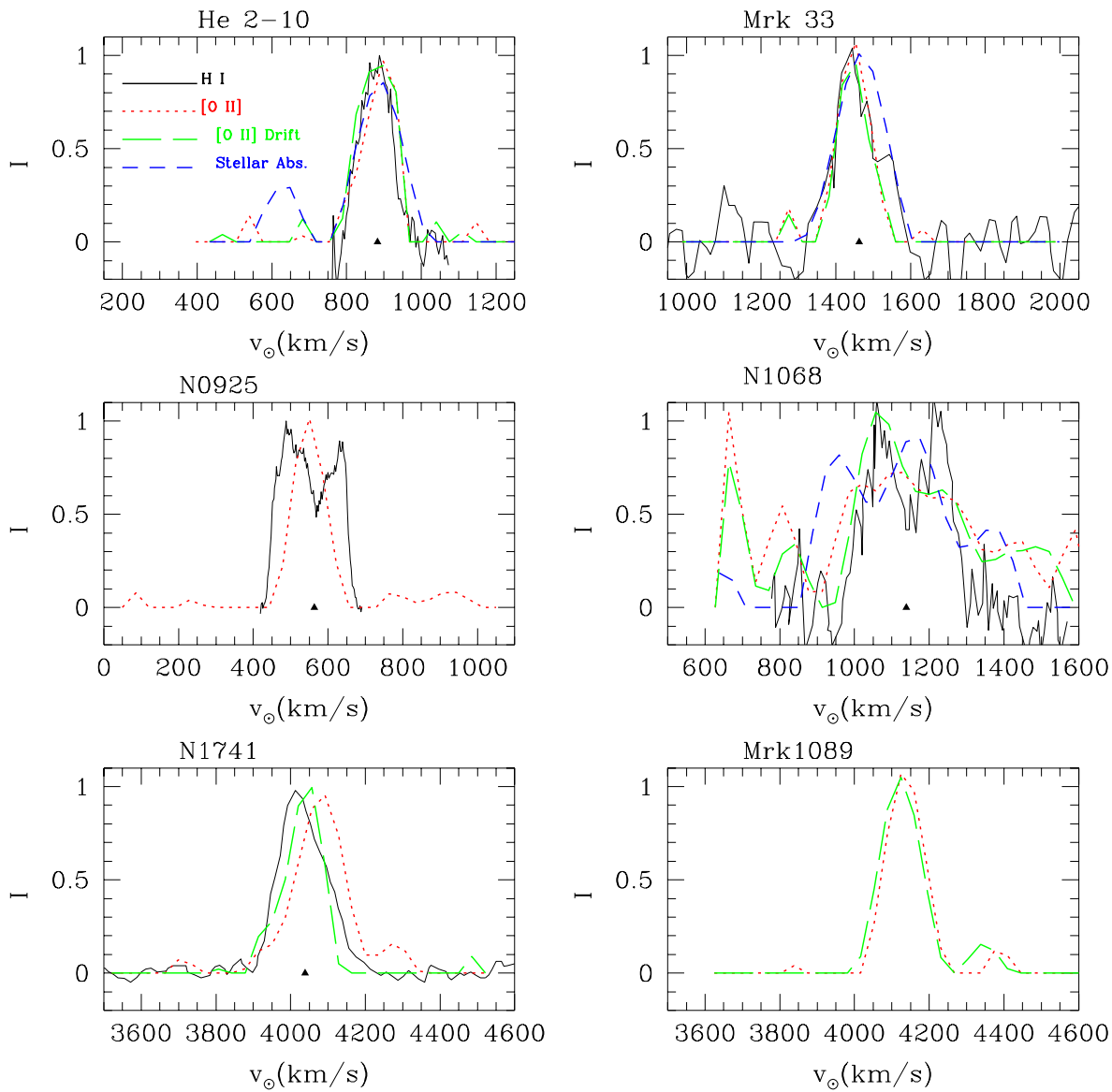}}
\figcaption[fig4] {Comparison of the velocity profiles for each galaxy from
neutral hydrogen (solid lines), [O~II] emission (dotted lines--fixed
slits; long dashes--drift
scan), and stellar absorption (medium dashes).  Absolute velocities
are shifted to a common (heliocentric) reference frame and
are accurate to $\sim8$~\kms~RMS for the optical data.
Amplitude normalization is arbitrarily adjusted to allow
easy inspection of the velocity widths by eye. 
There is generally good agreement between all three dynamical
indicators both in the case of low mass galaxies 
showing Gaussian velocity profiles (e.g., He 2-10)
and massive spiral galaxies showing classical double-peaked
profiles.  A solid triangle marks the systemic velocity as measured from 
the \HI\ spectra.}
{ \label{HIOplot1.ps} } \end{figure}

\clearpage

\centerline{\psfig{file=fig3b.cps,width=6.5in,angle=0}}

\clearpage

\centerline{\psfig{file=fig3c.cps,width=6.5in,angle=0}}

\clearpage

\centerline{\psfig{file=fig3d.cps,width=6.5in,angle=0}}

\clearpage

\begin{figure} 
\centerline{\psfig{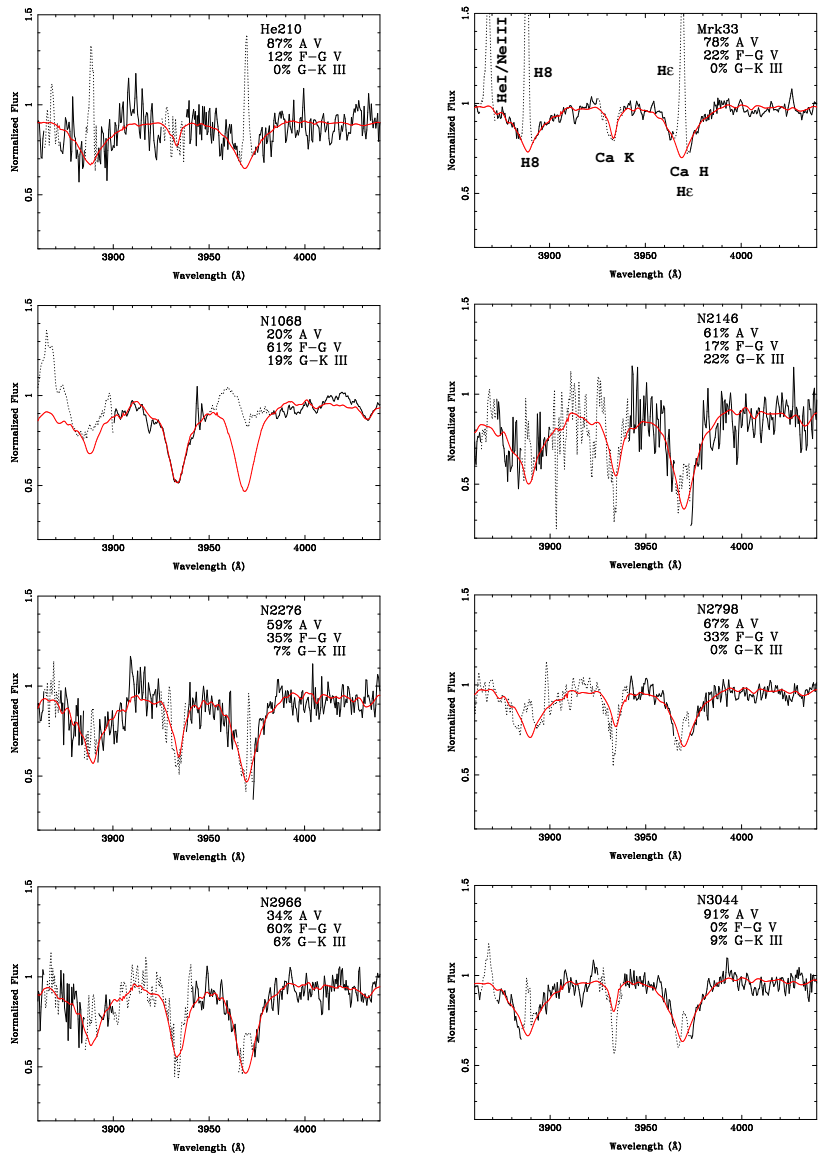}}
\figcaption[fig3] 
{Spectral region between 3850 \AA\ and 4040 \AA\ including 
several emission and absorption features:
blended [Ne III] $\lambda$3868/He I $\lambda$3867 in emission, 
He I $\lambda$3889 and $H8$ in emission and broad absorption near
$\lambda$3889, Ca H\& K absorption lines near $\lambda$3970
and $\lambda$3932, and the H$\epsilon$ seen in emission
and broad absorption at $\lambda$3970.  
The solid line overplotted on the raw data
shows the best fit synthesis of template stars
convolved with the measured velocity profile.  
Dashed lines show regions of the spectrum
contaminated by emission features excised for fitting purposes.
The label in each figure lists the fractional contribution from
A dwarfs, F--G dwarfs, and G-K giants in the best
fit composite spectrum. Note how the fraction of A stars
correlates with emission line strength.
{ \label{abspec} } }
\end{figure}

\clearpage

\centerline{\psfig{file=fig4b.cps,width=5.2in,angle=0}}

\clearpage

\begin{figure} \centerline{\psfig{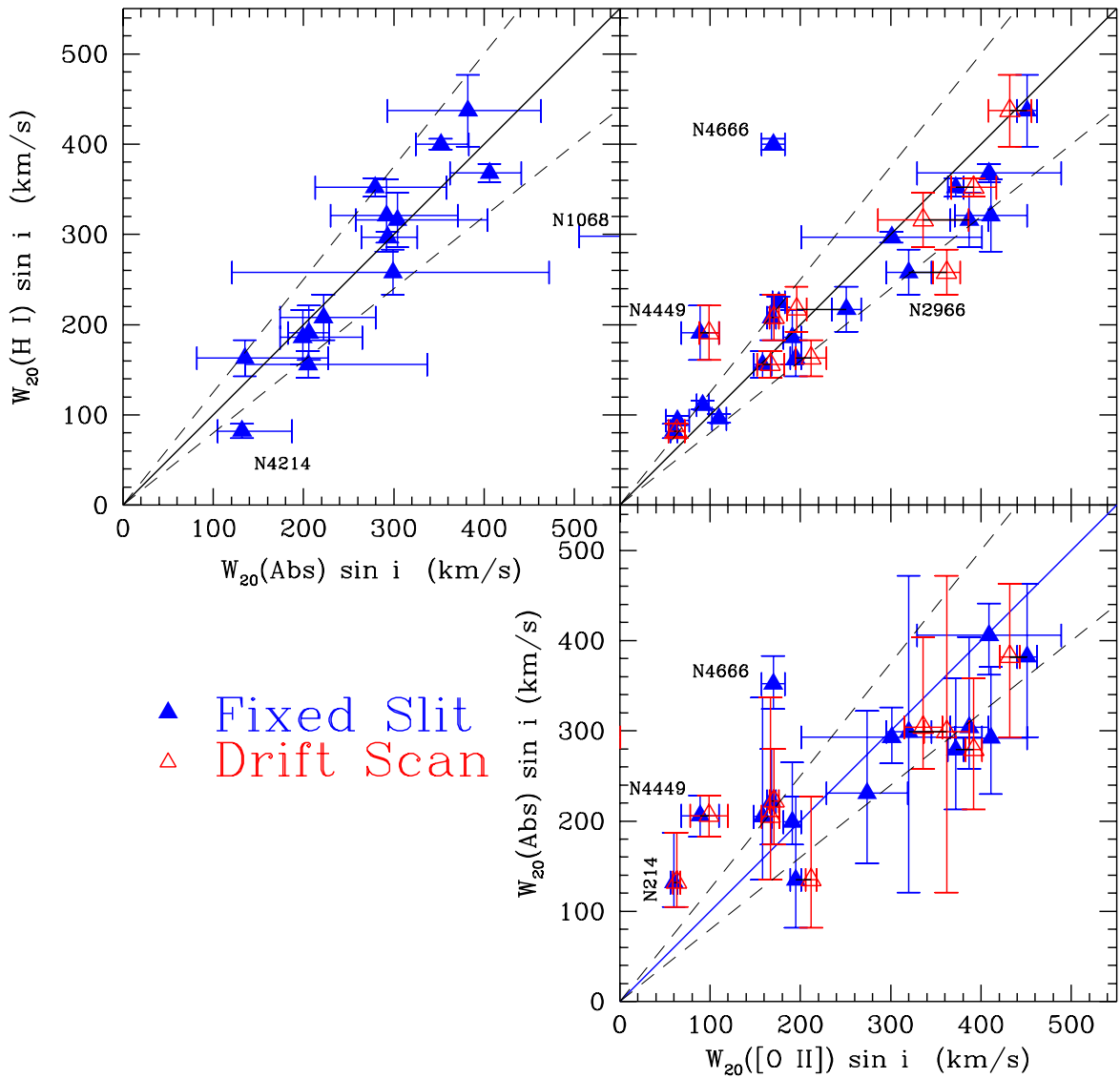}}
\figcaption[fig5] {Comparison of the velocity widths measured from
\HI, [O~II], and stellar absorption lines.  Dashed lines
indicate the 20\% variations from a 1:1 correspondence (solid line).
{\it Upper Left:} 
$W_{20}(HI)$ versus $W_{20}(Abs)$.  For most galaxies
both dynamical tracers yield comparable velocities within
20\% of each other.
{\it Upper Right:} $W_{20}(HI)$ versus $W_{20}([O~II])$.
Here, open symbols distinguish drift scan exposures from the
fixed slit exposures (solid symbols). 
The majority of the galaxies yield comparable velocity widths,
within 20\%, from both kinematic indicators.
{\it Lower Right:}  $W_{20}(Abs)$ versus $W_{20}([O~II])$. }
{ \label{vels.ps} } \end{figure}

\begin{figure} \centerline{\psfig{file=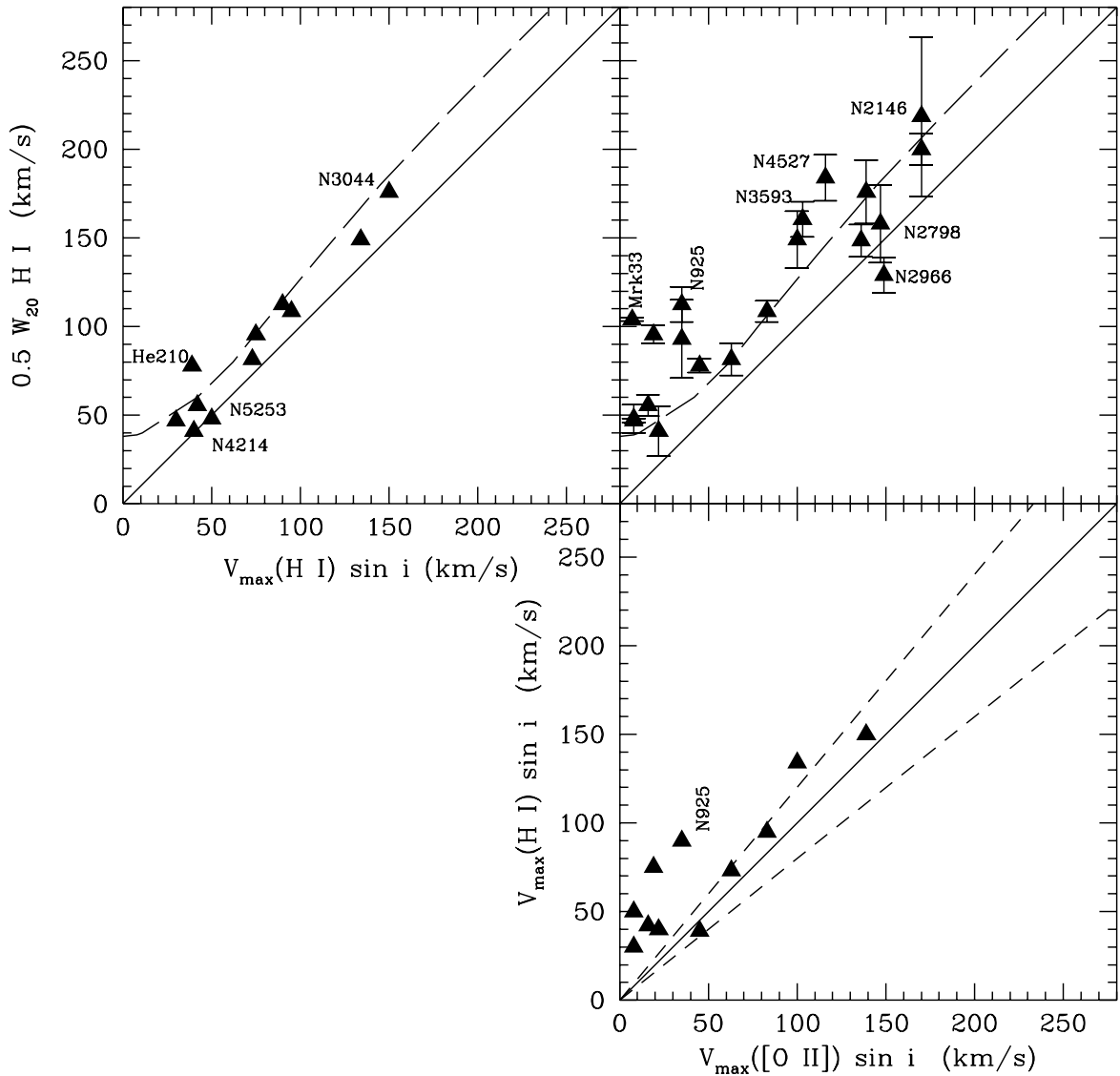,width=6.5in,angle=0}}
\figcaption[fig6] {Comparison of the rotation curves
as measured from
\HI\ 20\% width, \HI\ spatially-resolved
aperture synthesis data, the [O~II] rotation curves.
The solid line shows the 1:1 correspondence, while the
long-dashed line shows the relation between $W_{20}$ and
the maximum rotational amplitude advocated by Tully \& Fouqu\'e
(1985).  {\it Upper Left:} 
$W_{20}(HI)$ versus $W_{max}(HI)$.  
{\it Upper Right:} $W_{20}(HI)$ versus $V_{max}([O~II])$.
{\it Lower Right:}  $V_{max}(HI)$ versus $V_{max}([O~II])$.
A dashed line shows the 20\% deviations from 1:1.  In most cases, the [O~II]
rotation curve cannot be traced as far as the \HI\ rotation curve, leading to lower
$V_{max}$ estimates.}
{ \label{WHI.ps} } \end{figure}

\begin{figure} \centerline{\psfig{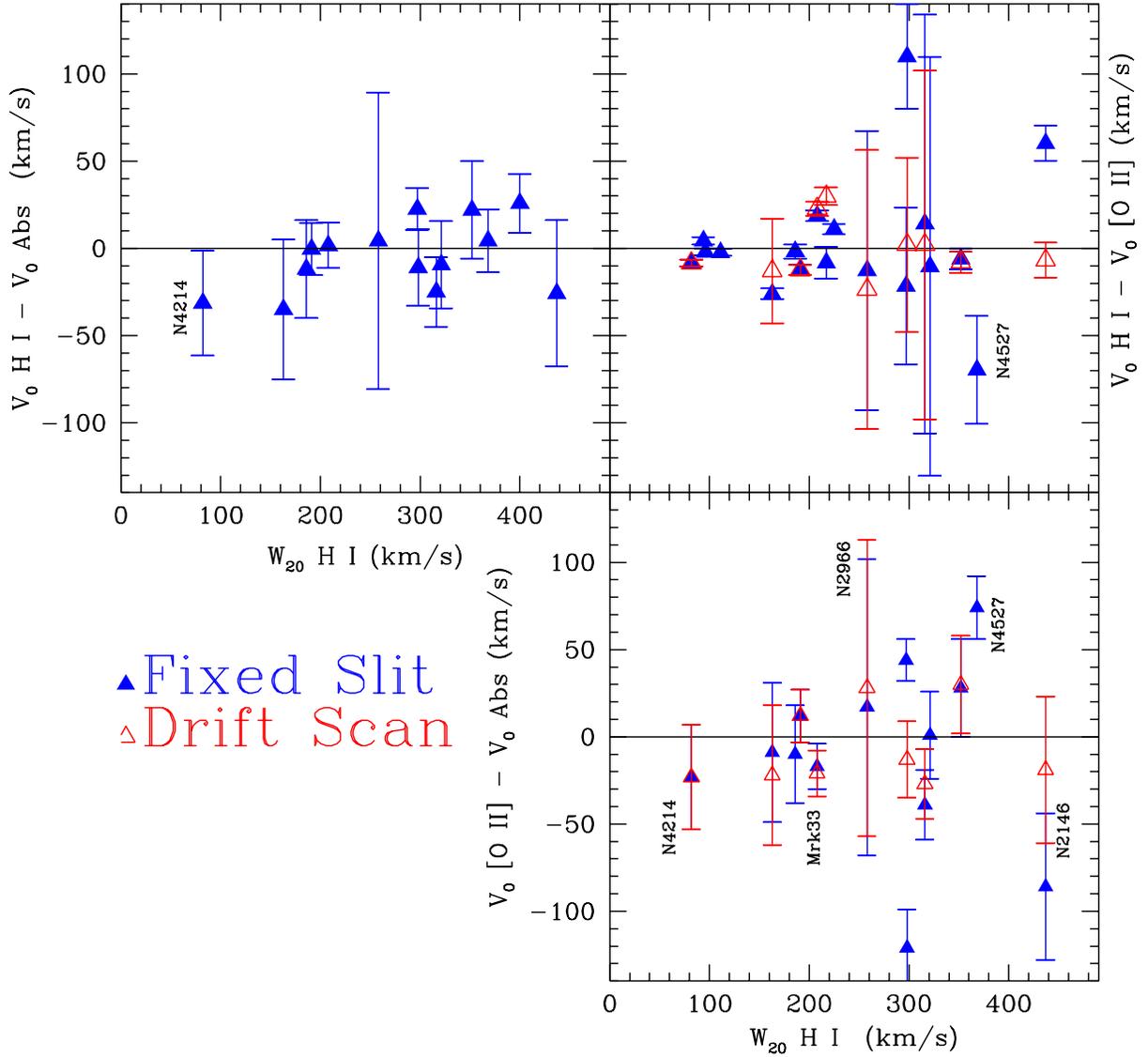}}
\figcaption[fig7] {Comparison of the differences between systemic
velocities as measured from
\HI, [O~II], and stellar absorption lines as a function of \HI\ width.
 {\it Upper Left:} $V_0(H~I) - V_0(Abs)$:   
{\it Upper Right:} $V_0(H~I) - V_0([O~II])$:
{\it Lower Right:}  $V_0([O~II]) - V_0(Abs)$:
While a few galaxies show evidence for real
deviation, much of the scatter is consistent with
measurement uncertainties.  In general, the drift scan
observations show better agreement than the fixed slits.
Most objects are consistent with
only small offsets of $\sim10$ \kms\ between the neutral gas,
ionized gas and stellar components.  
{ \label{V0b.ps} } }
\end{figure}

\end{document}


\begin{deluxetable}{llrrrrrrrrr}
\tablewidth{0pt}
\tablecaption{Observed Galaxy Sample  \label{Obs} }
\tablehead{
\colhead{Name1}  & 
\colhead{Name2}  & 
\colhead{Type}  & 
\colhead{Vel}   & 
\colhead{D}     & 
\colhead{$B-V$}   & 
\colhead{$M_B$}   & 
\colhead{Incl.} & 
\colhead{PA}    & 
\colhead{Drift} & \\ [.2ex]
\colhead{} &  
\colhead{} &  
\colhead{} &  
\colhead{km/s} &  
\colhead{(Mpc)} &  
\colhead{(mag)} &  
\colhead{(mag)} &  
\colhead{\degr} &  
\colhead{\degr} &  
\colhead{\arcsec} \\ [.2ex]
\colhead{(1)} &  
\colhead{(2)} &  
\colhead{(3)} &  
\colhead{(4)} &  
\colhead{(5)} &  
\colhead{(6)} &  
\colhead{(7)} &
\colhead{(8)} &
\colhead{(9)} &
\colhead{(10)} }
\startdata
He 2-10  & ESO 495-G021& I0 pec       &  873 & 10.5 & 0.33   & -19.0  & 32 & 135 & 10   \nl
Mrk 33   & UGC 5720    & Im pec       & 1461 & 24.9 & ... & -18.48 & 28 & 120 &  8   \nl
NGC 925  & UGC 1913    & SAB(s)d      &  553 & 9.4  & 0.57 & -19.66 & 54 & 102 & ...   \nl
NGC 1068 & UGC 2188    & SA(rs)b Sy2  & 1136 & 14.4 & 0.74 & -21.39 & 40 &  70 & 14   \nl
NGC 1741 & HCG 31      & Sm           & 4107 & 58.  & ...  & -21.18 & 62 &  70 & 20   \nl
Mrk 1089 & HCG 31      & Sm           & 4069 & 58.  & ...  & -20.51 & 80 &  70 & 20   \nl
NGC 2146 & UGC 3429    & SB(s)ab pec  &  893 & 17.2 & 0.79 & -20.24 & 36 & 148 & 30   \nl
NGC 2276 & UGC 3740    & SAB(rs)c     & 2410 & 36.8 & 0.52 & -21.19 & 29 &  20 & ...   \nl
NGC 2798 & UGC 4905    & SB(s)a pec   & 1726 & 27.1 & 0.72 & -19.67 & 75 & 160 & 20   \nl
NGC 2966 & UGC 5181    & SB?          & 2044 & 29.  & ...  & -19.0 & 70 &  70 & 12  \nl
NGC 3044 & UGC 5311    & SB(s)c       & 1292 & 20.6 & 0.53 & -19.74 & 90 & 115 & 10   \nl
NGC 3593 & UGC 6272    & SA(s)a       &  628 &  5.5 & 0.94 & -17.01 & 69 &  90 & 30   \nl
NGC 4214 & UGC 7278    & IAB(s)m      &  291 &  3.4 & 0.46 & -17.57 & 37 & 130 & 20   \nl
NGC 4218 & Haro 28     & Sa           &  729 & 25.3 & ...  & -18.82 & 60 & 142 & 14   \nl
NGC 4449 & UGC 7592    & IBm          &  207 &  3.0 & 0.41 & -17.66 & 43 &  45 & 12   \nl
NGC 4527 & UGC 7721    & SAB(s)bc     & 1736 & 13.5 & 0.86 & -19.72 & 68 &  67 & 12   \nl
NGC 4666 & UGC 7926    & SABc liner   & 1520 & 14.1 & 0.75 & -19.77 & 76 &  42 & ...   \nl
NGC 4818 & Mrk 9022    & SAB(rs)ab pec& 1155 & 21.5 & 0.89 & -20.66 & 78 &  10 & ...   \nl
NGC 5248 & UGC 8616    & SAB(rs)bc    & 1153 & 22.7 & 0.65 & -21.07 & 45 & 130 & 40   \nl
NGC 5253 & UGCA 369    & Im pec       &  404 &  3.2 & 0.43 & -16.66 & 77 &  45 & ...   \nl
UM 439   & UGC 6578    & pec, HII     & 1104 & 14.6 & 0.28  & -15.8  & 60 & 155 & ...   \nl
UM 462   & UGC 6850    & pec, HII     & 1012 & 13.9 & 0.24  & -16.1  & 50 &  70 & ...   \nl
\enddata 
\tablenotetext{1}{ Galaxy Name}
\tablenotetext{2}{ Alternate Name,} 
\tablenotetext{3}{ Morphological type, }
\tablenotetext{4}{ Systemic velocity from
the NED (NASA/IPAC Extragalactic Database)}
\tablenotetext{5}{ Distance in Mpc, as listed in Tully (1988) who use $H_0$=70 \kms\ 
	Mpc$^{-1}$ }
\tablenotetext{6}{Observed B-V color from RC3 except for UM 439 and UM 462 which
	are from Salzer \etal\ (1989)}
\tablenotetext{7}{ Absolute magnitude from Tully (1988) or computed from the assumed distance
	and apparent magnitude for objects not listed in the Catalog of Nearby Galaxies}
\tablenotetext{8}{Inclination to the line of sight from Tully (1988), except for
UM~439 and UM~462 which come from van Zee, Skillman, \& Salzer
	(1998) and NGC~1741, NGC~2966 and Mrk 1089 which are from computed from the
	axial ratios, $R_{25}$, of RC3 using the Hubble-Holmberg relation $cos^2i=(R_{25}-0.04)/(0.96)$ . } 
\tablenotetext{9}{Position angle of the longslit used in the spectroscopic observations reported here}
\tablenotetext{10}{Full amplitude of the longslit drift scan in arcseconds}
\end{deluxetable}
\clearpage

\begin{deluxetable}{llrrr}
\tablewidth{0pt}
\tablecaption{Stars Used for Template Spectra \label{Obs} }
\tablehead{
\colhead{Name}  & 
\colhead{Type}  & 
\colhead{$T_{eff}$}  & 
\colhead{log g}  & 
\colhead{[Fe/H]}  }
\startdata
HD12929  & K2III & 4461.  &  2.40  & -0.03  \nl
HD27697  & K0III & 4975.  &  2.14  &  0.12  \nl
HD70110  & F9V   & 5867.  &  4.07  &  0.07  \nl
HD110014 & K2III & 4285    & 2.05   & 0.10   \nl
HD113226 & G8III & ...    &  2.10  &  0.00  \nl
HD114710 & G0V   & 5927.  &  4.38  &  0.03   \nl
HD124897 & K1III & 4276.  &  2.10  & -0.42  \nl
HD135722 & G8III & 4834.  &  2.45  & -0.44  \nl
HD140573 & K2III & 4579.  &  2.14  &  0.35  \nl
HD142860 & F6V   & 6175.  &  4.25  & -0.16  \nl
HD65900  & A1V   & 9692.  &  4.00  &  0.24  \nl
HD107259 & A2IV  & 9333.  &  3.00  &  0.11  \nl
\enddata 
\end{deluxetable}

\clearpage

\begin{deluxetable}{llrrrrrrrrrr}
\tablewidth{0pt}
\tablecaption{Kinematic Properties of the Sample  \label{Obs} }
\tablehead{
\colhead{Name}  & 
\colhead{HI}  & 
\colhead{HI}   & 
\colhead{HI}     & 
\colhead{HI}   & 
\colhead{HI}   & 
\colhead{[O~II]} & 
\colhead{[O~II]}    & 
\colhead{[O~II]} & 
\colhead{Abs} & 
\colhead{Abs} \\ [.2ex]
\colhead{} &  
\colhead{$V_0$} &  
\colhead{$W_{20}$} &  
\colhead{Ref.} &  
\colhead{$V_{max}$} &  
\colhead{Ref.} &  
\colhead{$V_0$} &  
\colhead{$W_{20}$} &  
\colhead{$V_{max}$} &  
\colhead{$V_0$} &  
\colhead{$W_{20}$} \\ [.2ex]
\colhead{(1)} &  
\colhead{(2)} &  
\colhead{(3)} &  
\colhead{(4)} &  
\colhead{(5)} &  
\colhead{(6)} &  
\colhead{(7)} &
\colhead{(8)} &
\colhead{(9)} &
\colhead{(10)} &
\colhead{(11)} } 
\startdata
He 2-10  & 873  & 156$\pm15$ &  A & 39 & A &  853$\pm$ 6 & 158$\pm$10 & 45$\pm$4 &  859$\pm$61 & 205$^{+132}_{- 70}$ \nl
Mrk 33   & 1467 & 208$\pm25$ &  B &    &   & 1413$\pm$ 3 & 168$\pm$ 5 &  8$\pm$1 & 1440$\pm$13 & 187$^{+ 58}_{- 48}$ \nl
NGC 925  & 551  & 222$\pm$6  &  C & 118 & D &  552$\pm$ 3 & 176$\pm$ 7 & 35$\pm$10 & ...         & ... \nl
NGC 1068 & 1137 & 298$\pm$30 &  E &134 & F & 1029$\pm$30 & 1129$\pm$200& 100$\pm$16& 1160$\pm$22& 564$^{+ 38}_{- 59}$ \nl
NGC 1741 & 4047 & 217$\pm25$ &  G & 95 & H & 4049$\pm$ 9 & 251$\pm$16 & 83$\pm$6 & ...         & ... \nl
Mrk 1089 & ...  & ...        &  G & 95 & H & 4118$\pm$ 3 & 188$\pm$ 9 & 25$\pm$3 & ...         & ... \nl
NGC 2146 & 903  & 437$\pm$40 &  I &    &   &  854$\pm$10 & 451$\pm$11 & 170$\pm$45&  949$\pm$42 & 382$^{+ 81}_{- 89}$ \nl
NGC 2276 & 2419 & 186$\pm$30 &  J &    &   & 2417$\pm$ 4 & 191$\pm$10 &  35$\pm$22& 2427$\pm$28 & 199$^{+ 66}_{- 25}$ \nl
NGC 2798 & 1744 & 316$\pm$30 &  K &    &   & 1734$\pm$120& 387$\pm$21 & 147$\pm$22& 1756$\pm$20 & 304$^{+100}_{- 46}$ \nl
NGC 2966 & 2042 & 258$\pm25$ &  L &    &   & 2044$\pm$80 & 320$\pm$25 & 150$\pm$10& 2027$\pm$85 & 299$^{+173}_{-178}$ \nl
NGC 3044 & 1298 & 352$\pm$10 &  M & 150& M & 1291$\pm$ 6 & 372$\pm$ 9 & 139$\pm$18& 1273$\pm$28 & 279$^{+ 79}_{- 66}$ \nl
NGC 3593 & 651  & 321$\pm$40 &  N &    &   &  661$\pm$121& 411$\pm$40 & 103$\pm$10&  660$\pm$25 & 292$^{+ 79}_{- 62}$ \nl
NGC 4214 & 293  &  82$\pm$8  &  I & 40 & O &  300$\pm$ 2 &  60$\pm$ 4 & 22$\pm$14 &  313$\pm$30 & 132$^{+ 55}_{- 27}$ \nl
NGC 4218 & 724  & 163$\pm$20 &  B & 73 & P &  750$\pm$ 3 & 195$\pm$ 6 & 62$\pm$9 &  759$\pm$40 & 135$^{+ 92}_{- 53}$ \nl
NGC 4449 & 214  & 191$\pm$30 &  Q & 75 & Q &  214$\pm$ 3 &  89$\pm$21 & 19$\pm$5 &  203$\pm$15 & 135$^{+ 22}_{- 23}$ \nl
NGC 4527 & 1736 & 368$\pm$10 &  R &    &   & 1794$\pm$31 & 409$\pm$80 &116$\pm$13 & 1730$\pm$18 & 406$^{+ 35}_{- 44}$ \nl
NGC 4666 & 1523 & 400$\pm$6  &  S &    &   & 1685$\pm$ 8 & 170$\pm$13 &170$\pm$9 & 1495$\pm$17 & 352$^{+ 31}_{- 28}$ \nl
NGC 4818 & ...  & ...        &  T &    &   &  984$\pm$20 & 274$\pm$45 & 21$\pm$9 &  969$\pm$30 & 231$^{+ 91}_{- 78}$ \nl
NGC 5248 & 1153 & 297$\pm$6  &  U &    &   & 1160$\pm$45 & 301$\pm$100&136$\pm$9 & 1138$\pm$12 & 293$^{+ 33}_{- 29}$ \nl
NGC 5253 &  405 &  96$\pm$20 &  V & 50 & W &  401$\pm$ 2 & 110$\pm$ 8 &8$\pm$8  & ...         & ... \nl
UM 439   & 1097 & 111$\pm 5$ &  X & 42 & X & 1069$\pm$ 2 &  92$\pm$ 7 &16$\pm$6  & ...         & ... \nl
UM 462   & 1057 &  94$\pm 5$ &  X & 30 & X & 1053$\pm$ 2 &  64$\pm$13 & 8$\pm$8  & ...         & ... \nl
\enddata 
\tablerefs{
(A) Kobulnicky \etal\ 1995
(B) Gordon \& Gottesman 1981
(C) Pisano, Wilcots, \& Elmegreen 1998
(D) Gottesman 1980
(E) Staveley-Smith \& Davies 1987
(F) Brinks \etal\ 1997
(G) Williams \& Rood 1987
(H) Williams, McMahon, \& van Gorkom 1991
(I) Tifte \& Cocke  1988
(J) Peterson 1979
(K) Huchtmeier 1982
(L) Lewis, Helou \& Salpeter 1985
(M) Lee \& Irwin 1997
(N) Huchtmeier \& Seiradakis 1985
(O) Kobulnicky 1999, private communication 
(P) Verheijen 1997
(Q) Bajaja, Huchtmeier \& Klein 1994
(R) Hoffman \etal\ 1989
(S) Richter \& Huchtmeier 1987
(T) Theureau \etal\ 1998
(U) Shostak 1978
(V) Kobulnicky, unpublished Parkes 64~m observation
(W) Kobulnicky \& Skillman 1995
(X) Van Zee, Skillman, \& Salzer 1998
}
\tablenotetext{1}{Galaxy Name}
\tablenotetext{2}{21-cm systemic velocity as re-measured by us from
digitized spectra presented in the original reference.} 
\tablenotetext{3}{Observed velocity width at the 20\% height of the 21-cm spectrum 
	as cited in the reference in Column 4.  The associated uncertainty,
	if listed, is estimated from the 
	dispersion in the 20\% widths reported by different authors as tabulated
	in Huchtmeier \& Richter (1988).}
\tablenotetext{4} {Reference for 21-cm spectra. }
\tablenotetext{5} {Maximum projected neutral hydrogen rotational velocity,
	$V_{max}$, relative to the systemic velocity, from aperture synthesis 
	observations, where such data exist.}
\tablenotetext{6}{Reference for 21-cm aperture synthesis HI data.}
\tablenotetext{7}{Systemic velocity computed from the
	deconvolved [O~II]  profiles.  We adopt the midpoint of the 
	20\% peak. Uncertainties are {\it relative} values only.  An additional
	uncertainty of 8 km/s must be added in quadrature to these value to
	obtain errors on the absolute radial velocities in the heliocentric frame of reference. }
\tablenotetext{8} {Measured velocity width at the 20\% height of the 
	deconvolved [O~II] spectra.  Here, we cite the values measured from 
	the drift scan exposures, where available.}
\tablenotetext{9} {Maximum projected rotational velocity as measured
	from the [O~II] rotation curves.}
\tablenotetext{10} {Systemic velocity computed from the
	deconvolved stellar absorption profiles.  We adopt 
	the midpoint of the 20\% peak. }
\tablenotetext{11} {Measured velocity width at the 20\% height of the 
	deconvolved absorption spectra.  The fixed slit exposures,
rather than drift scans are used since they have higher S/N in
the stellar continuum.}
\end{deluxetable}